\documentclass[useAMS,usenatbib]{mn2e}
\usepackage{hyperref}
\hypersetup{
     colorlinks   = true,
     citecolor    = blue,
     linkcolor     = blue
}
\usepackage{caption}

\usepackage[T1]{fontenc}
\usepackage{ae,aecompl}
\usepackage{natbib}
\def\fesc{\ifmmode f_{\rm esc} \else $f_{\rm esc}$\fi}

\usepackage{graphicx}	
\usepackage{amsmath}	
\usepackage{amssymb}	

\def\apj{ApJ}
\def\apjs{ApJS}

\def\aap{A\&A}
\def\aaps{A\&AS}

\def\mnras{MNRAS}

\def\pasp{PASP}
\def\nat{Nature}

\def\araa{Ann.Rev.Astron.Astrophys.}

\title[The bursting nature of star formation]{The bursting nature
of star formation in compact star-forming galaxies from the Sloan Digital Sky Survey}

\author[Y. I. Izotov et al.]{
Y. I.\ Izotov$^{1,2}$,\thanks{E-mail: izotov@mao.kiev.ua}
N. G.\ Guseva$^{1,2}$, 
K. J.\ Fricke$^{2,3}$
\& C.\ Henkel$^{2,4}$
\\
$^{1}$Main Astronomical Observatory, Ukrainian National Academy of Sciences,
27 Zabolotnoho str., Kyiv 03680, Ukraine\\
$^{2}$Max-Planck-Institut f\"ur Radioastronomie, 
                     Auf dem H\"ugel 
                     69, 53121 Bonn, Germany\\
$^{3}$Institut f\"ur Astrophysik, 
                     G\"ottingen Universit\"at, Friedrich-Hund-Platz 1, 
                     37077 G\"ottingen, Germany\\
$^{4}$Astronomy Department, King Abdulaziz University, 
                     P.O. Box 80203, Jeddah 21589, Saudi Arabia
}

\date{Accepted XXX. Received YYY; in original form ZZZ}

\pubyear{2015}

\begin{document}
\label{firstpage}
\pagerange{\pageref{firstpage}--\pageref{lastpage}}
\maketitle

\begin{abstract}
We study integrated characteristics of $\sim$ 14000 low-redshift 
(0 $<$ $z$ $<$ 1) compact 
star-forming galaxies (SFGs) selected from the Data Release 12 of the Sloan 
Digital Sky Survey. 
It is found that emission of these galaxies is dominated by strong young bursts 
of star formation, implying that their luminosities experience
rapid variations on a time scale of a few Myr. 
Reducing integrated characteristics of these galaxies
to zero burst age would result in a considerably tighter and almost linear 
relation between stellar mass and star formation rate (SFR). The same 
correction implies that the specific star formation rate (the ratio of SFR and
stellar mass) is not dependent on the 
galaxy stellar mass. We conclude that the correction for rapid luminosity 
evolution must be taken into account in a similar way when comparing different 
samples of low- and high-redshift SFGs. If the bursting 
nature of star formation and young burst ages are characteristics of the 
galaxies selected at high redshifts, the age correction of observed SFRs 
derived from the H$\beta$ emission line or UV continua would modify
the derived SFR densities in the early universe.
\end{abstract}




\section{Introduction}\label{sec1}

Relations between luminosities, stellar masses, and star formation rates (SFRs)
of star-forming galaxies (SFGs) and their variations with redshift play an 
important role in our understanding of their formation and evolution
on cosmological time scales.

These relations for various samples of low-$z$ and high-$z$ 
galaxies were studied in numerous papers
\citep[e.g. ][]{T04,M08,M10,S14,I11,I15}. 
For example, \citet{M08} found an evolution of
the stellar mass-metallicity relation with redshift in the sense
that high-$z$ ($z$ $\sim$ 3.5) galaxies
are more metal-poor compared to low-$z$ galaxies for the same stellar
mass. This effect is more pronounced for low-mass galaxies.
\citet{M10} considered a more general relation between stellar mass $M_\star$,
metallicity, and SFR. They found that high-$z$
galaxies are characterised by progressively higher SFRs.

On the other hand, there is some evidence that properties of 
low-$z$ ``Green Pea'' (GP) galaxies and Luminous Compact Galaxies (LCGs) and 
high-$z$ Lyman-alpha emitting (LAE) and Lyman-break 
galaxies (LBG) are similar with respect to 
the stellar mass, SFR and metallicity \citep{Ca09,I11,I15}.
It was found by \citet{I14a} and \citet{I15} that luminosity-metallicity, 
mass-metallicity and mass-SFR relations for Sloan Digital Sky Survey (SDSS) 
compact SFGs
are much flatter than e.g. the \citet{T04} relations for the main sequence 
low-$z$ SFGs.
They also found that low-$z$ galaxies with high SFRs do not deviate from
the relation established for galaxies with lower SFRs. Furthermore, no
considerable redshift dependence was found.


These results indicate inconsistencies between relations obtained
in different studies.
Among possible sources of inconsistencies are different criteria for the 
sample selection and a variety of strong-line methods, which are used for the 
metallicity determination. Element abundances derived by different methods can 
be different by more than 0.5 dex. 

Furthermore, it is generally assumed that star formation in SFGs 
occurs continuously at time scales of $\sim$100 Myr either with a 
constant or an exponentially decreasing SFR. However, the 
possibility that star formation may occur in strong bursts 
with rapidly changing H$\beta$ line and UV continuum luminosities by a factor of
several times on time scales of $\leq$10 Myr is rarely taken into account 
\citep{I14a,I15}. The presence of 
bursts may only slightly change the deduced total stellar masses, which are 
determined from the galaxy continuum emission in the visible and longer 
wavelength ranges, but may significantly affect the H$\beta$ emission line and 
UV luminosities and hence SFRs
derived from these luminosities. As a consequence, the comparison of
bursting SFGs in different samples should be done 
only after reduction of their luminosities to the same burst age 
\citep[e.g. ][]{I15}.

In this paper we consider an impact of star formation history
on the H$\beta$ line and UV continuum luminosities and present evidence that 
star formation in compact SFGs has a bursting nature. 
In Sect. \ref{sec2} we discuss the sample. The technique used for the 
determination of luminosities and stellar masses is described in 
Sect. \ref{sec3}. Evidence for the bursting nature of star formation in
SDSS compact SFGs is discussed in Sect. \ref{sec4}. The relations
stellar mass - SFR and stellar mass - specific SFR (sSFR) are considered in 
Sect. \ref{sec5}. The main results of the paper are summarised in 
Sect. \ref{sec6}.

\section{The sample of compact star-forming galaxies}\label{sec2}

\subsection{Selection criteria}

   The SDSS Data Release 12 (DR12)
\citep{A15} was used to select a sample of compact SFGs
applying selection criteria by \citet{I15}:
1) the angular galaxy radius on the SDSS images is $R_{50}$ $\leq$ 3 arcsec, 
where $R_{50}$ is the galaxy's Petrosian radius within which 50\% of the 
galaxy's flux in the SDSS $r$ band is contained;
2) H {\sc ii} regions in spiral galaxies were excluded using SDSS images; 
3) galaxies with AGN activity were excluded using line ratios. 
Applying these criteria, $\sim$ 14000 galaxies with redshifts 
$z$ $<$ 1 were selected.

\begin{figure}
\includegraphics[angle=-90,width=0.95\linewidth]{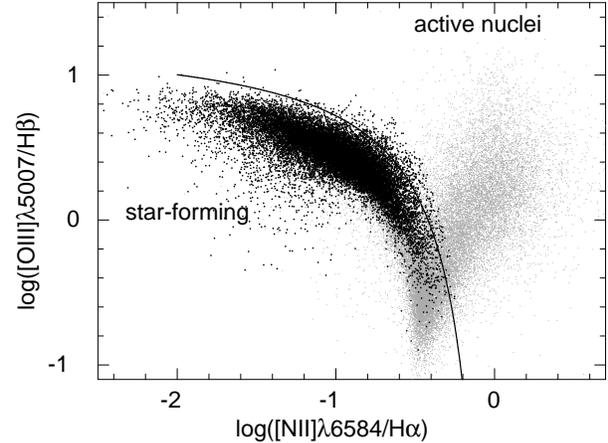}
\caption{The Baldwin-Phillips-Terlevich (BPT) 
diagram \citep{BPT81} for narrow emission-line galaxies.
The compact star-forming galaxies (this paper) are represented by black dots.
Also, plotted are emission-line galaxies from SDSS DR7 (cloud
of grey dots). The solid line by \citet{K03} separates 
star-forming galaxies (SFG) from active galactic nuclei (AGN).
\label{fig1}}
\end{figure}

The location of the selected galaxies in the 
[O~{\sc iii}]$\lambda$5007/H$\beta$ -- [N~{\sc ii}]$\lambda$6584/H$\alpha$ 
diagnostic diagram \citep*{BPT81} is shown in Fig. \ref{fig1}. The solid line 
by \citet{K03} separates SFGs and active galactic nuclei (AGN). Our sample
galaxies are located in the SFG region and thus their 
interstellar medium is ionized by hot stars in the star-forming regions.

\begin{figure}
\includegraphics[angle=-90,width=0.80\linewidth]{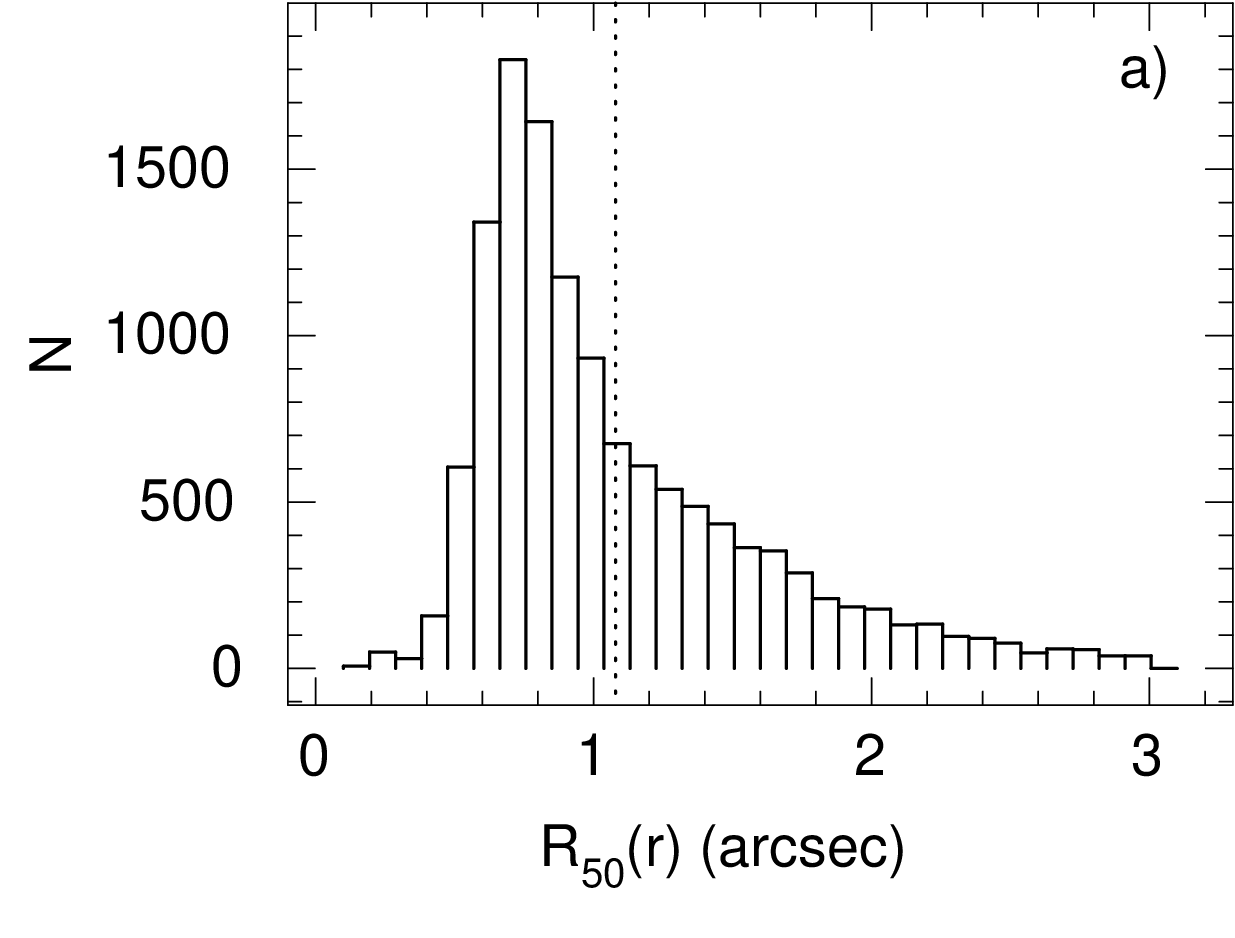}
\includegraphics[angle=-90,width=0.80\linewidth]{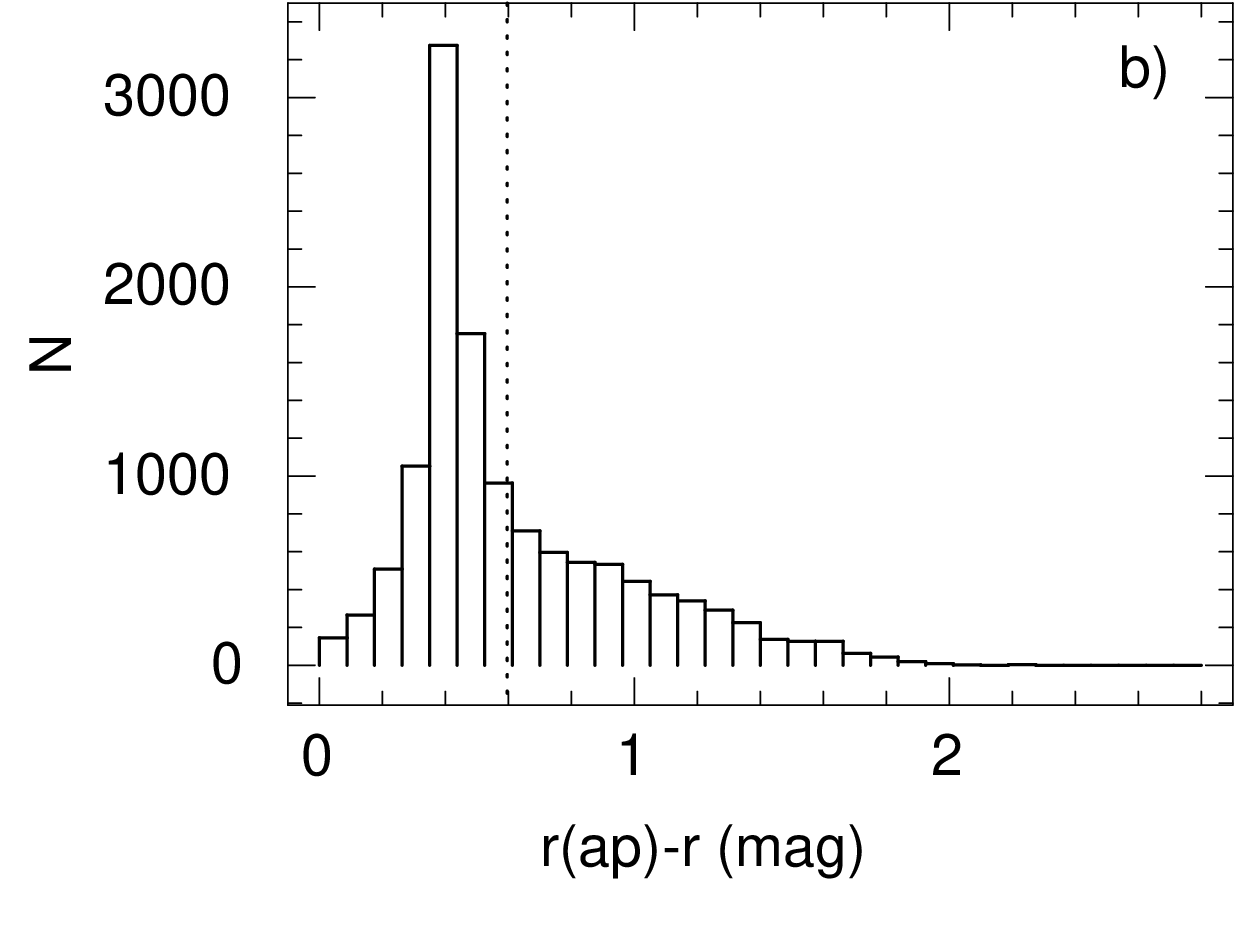}
\caption{Distributions in the SDSS $r$ band of (a) Petrosian radii 
$R_{50}$($r$) and (b) aperture corrections for our compact SFGs. 
Vertical dotted lines in both panels indicate average values.
\label{fig2}}
\end{figure}

The distribution of $R_{50}$ for our galaxies with averaged value
of $\sim$ 1 arcsec is presented in Fig. \ref{fig2}a. These small angular sizes
indicate that a considerable fraction of the galaxy light is concentrated
inside the spectroscopic aperture. This results in relatively
small aperture corrections with average value $\sim$ 0.6 mag to convert the
fluxes derived inside the spectroscopic aperture to the total galaxy fluxes
(Fig. \ref{fig2}b). Furthermore, small angular sizes also indicate that most
of the selected galaxies are unresolved or barely resolved on the SDSS images.
True angular sizes can be much smaller. This was demonstrated e.g. by
\citet{I16b} who found angular radii $R_{50}$ $\la$ 0.1 arcsec for 
five compact SFGs at redshifts $\sim$ 0.3. All this implies that our 
galaxies have small linear radii of a few kpc. We note that the adopted
upper limit $R_{50}$ = 3 arcsec is somewhat arbitrary. It is two and three
times higher than the radii of spectroscopic apertures used in the SDSS-II and
SDSS-III surveys, respectively. Reducing $R_{50}$ to the radii of spectroscopic 
apertures will not influence the selection of distant SFGs with $z$ $\ga$ 0.3, 
but many lower-redshift SFGs with similar properties will be missed. 
Therefore, for the purpose of this paper the criterion $R_{50}$ $\leq$ 3 arcsec
is optimal.

\begin{figure}
\includegraphics[angle=-90,width=0.95\linewidth]{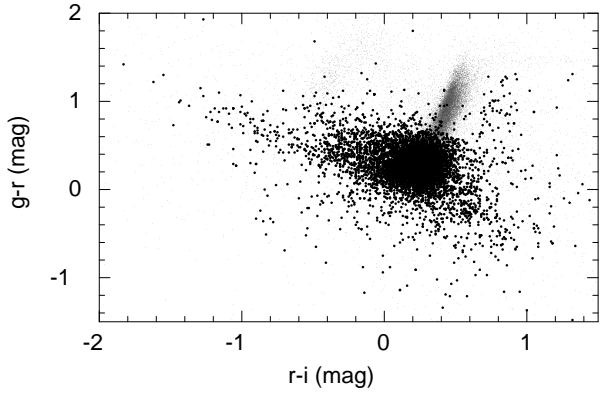}
\caption{$(g-r)-(r-i)$ colour-colour diagram for compact SFGs
(black symbols). For comparison, a representative sample of SDSS ``normal''
galaxies is shown by grey symbols.
\label{fig3}}
\end{figure}

The location of our selected SFGs on the SDSS $(g-r) - (r-i)$ colour-colour 
diagram is represented in Fig. \ref{fig3} by black symbols. This diagram was 
used by \citet{Ca09} to select GPs with strong emission lines
at redshifts $z$ $\sim$ 0.13 -- 0.3. They are located in the upper left part of 
the diagram in the region with $(r-i)$ $\la$ $-$0.5 mag $(g-r)$ $\ga$ 0.5 mag. 
Our compact SFGs are distributed in a similar way as the LCGs
 by \citet{I11} (their fig. 4). The galaxies
with high EW(H$\beta$) $\ga$ 100\AA\ are distributed in a very wide range
of the $(r - i)$ colours, which is determined by the observed wavelength
of the strong redshifted 
[O {\sc iii}]$\lambda$5007\AA\ emission line. Galaxies at redshifts of 
$\sim$ 0.2 -- 0.3 are located in the upper left corner of the diagram, while 
galaxies at $z$ $\sim$ 0.5 are in the lower right corner with $(r-i)$ 
$\sim$ 0.5 -- 1.0.
The distribution of our SFGs with low EW(H$\beta$) $<$ 50\AA\ is more compact
because of the much weaker [O {\sc iii}]$\lambda$5007\AA\ emission. Most of
these SFGs are located in the $(r-i)$ range 0.0 -- 0.5 and are coincident
with the location of the SDSS QSOs with blue continua, as it was shown by
\citet{I11}. On the other hand, compact SFGs are clearly different from the
SDSS ``normal'' quiescent galaxies (dense cloud of grey symbols with
$(r-i) \sim 0.4 - 0.6$ mag and $(g-r) > 0.6$ mag), which have
redder $(g-r)$ colour.

The SDSS photometric and spectroscopic data in combination with the 
{\sl Galaxy Evolution Explorer} ({\sl GALEX}) photometric data are used in
Section \ref{sec3} to derive galaxy stellar masses $M_\star$, H$\beta$ 
luminosities $L$(H$\beta$), far-UV (FUV) and near-UV (NUV) luminosities, SFRs 
and sSFRs. 

\begin{figure}
\includegraphics[angle=-90,width=0.84\linewidth]{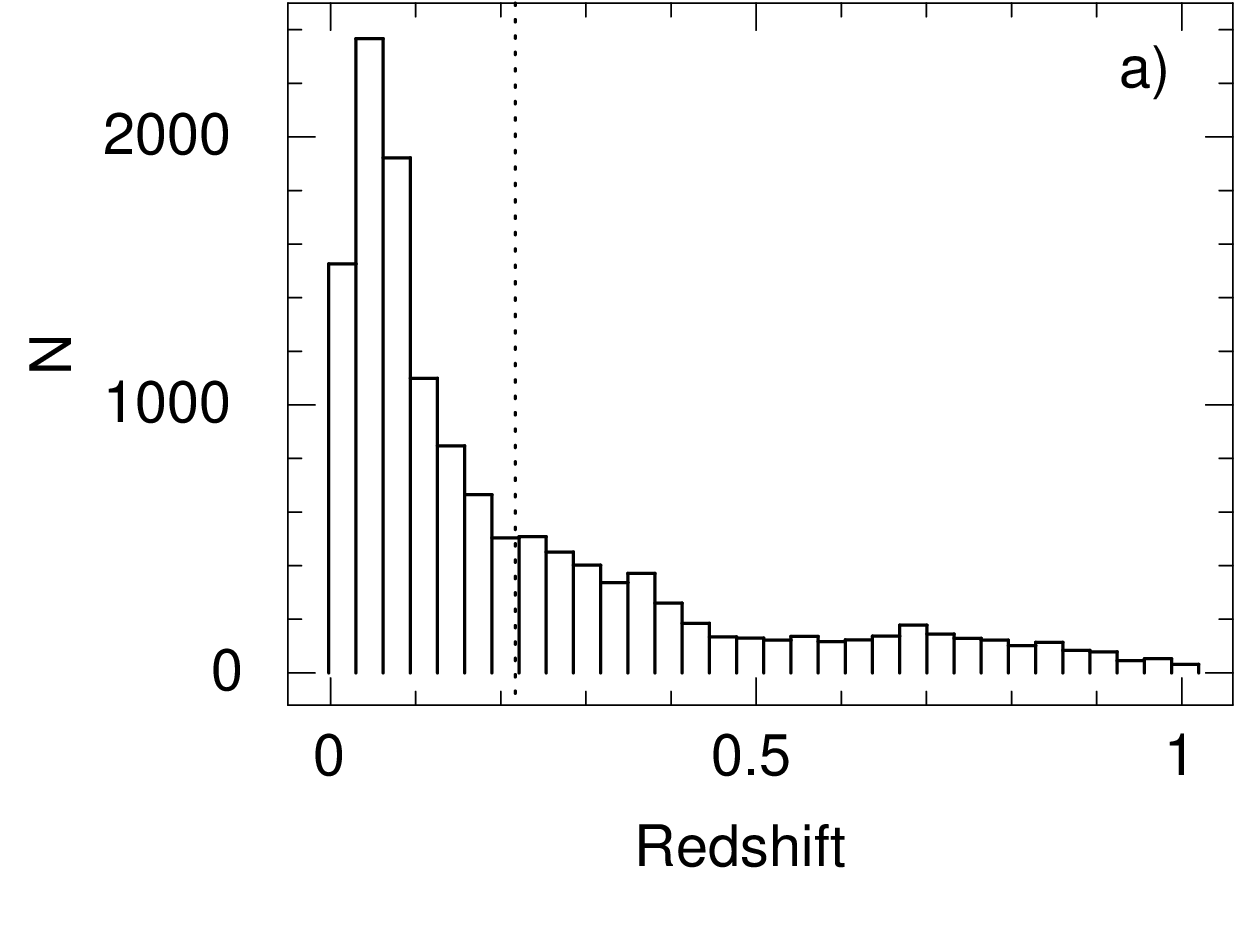}
\includegraphics[angle=-90,width=0.84\linewidth]{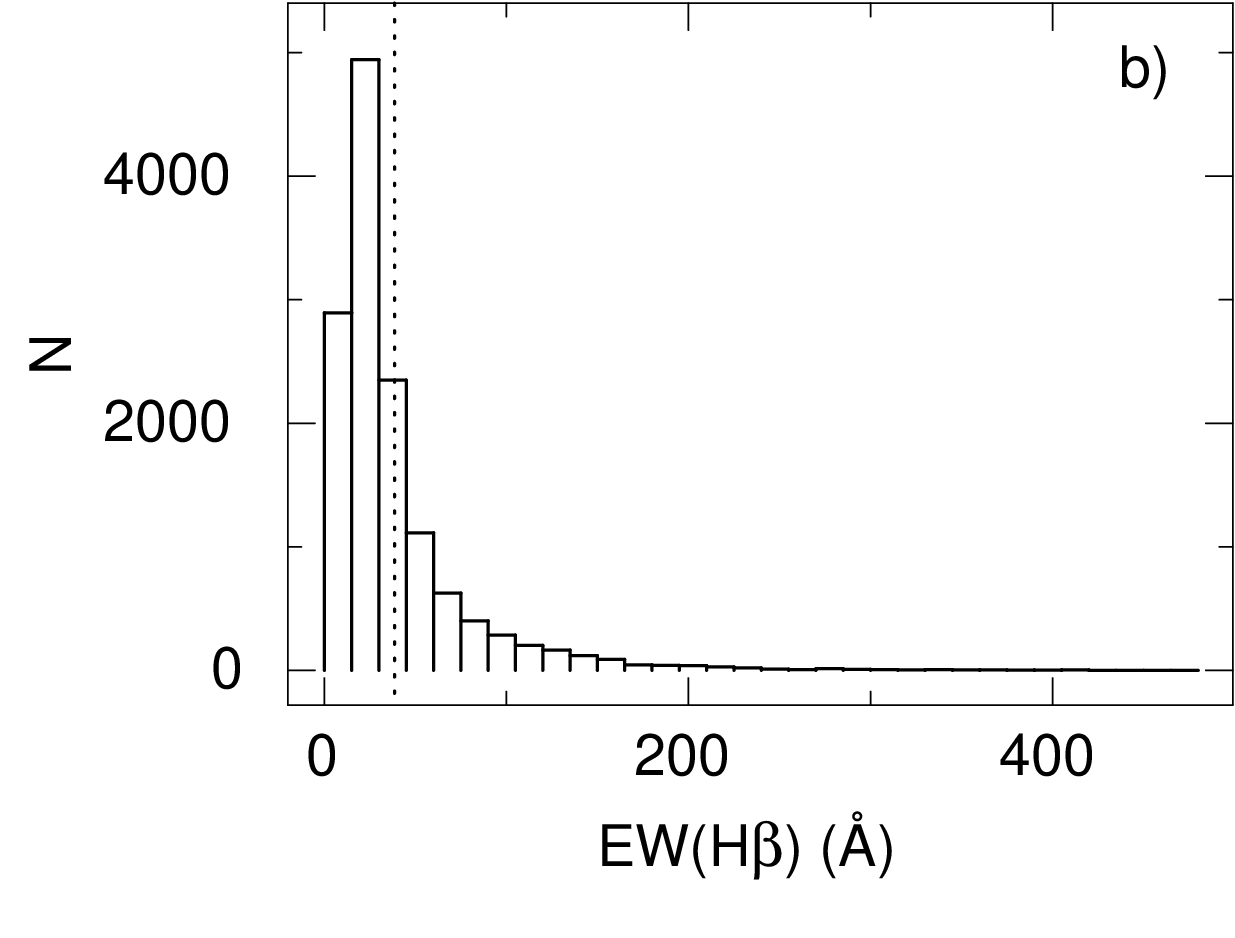}
\includegraphics[angle=-90,width=0.84\linewidth]{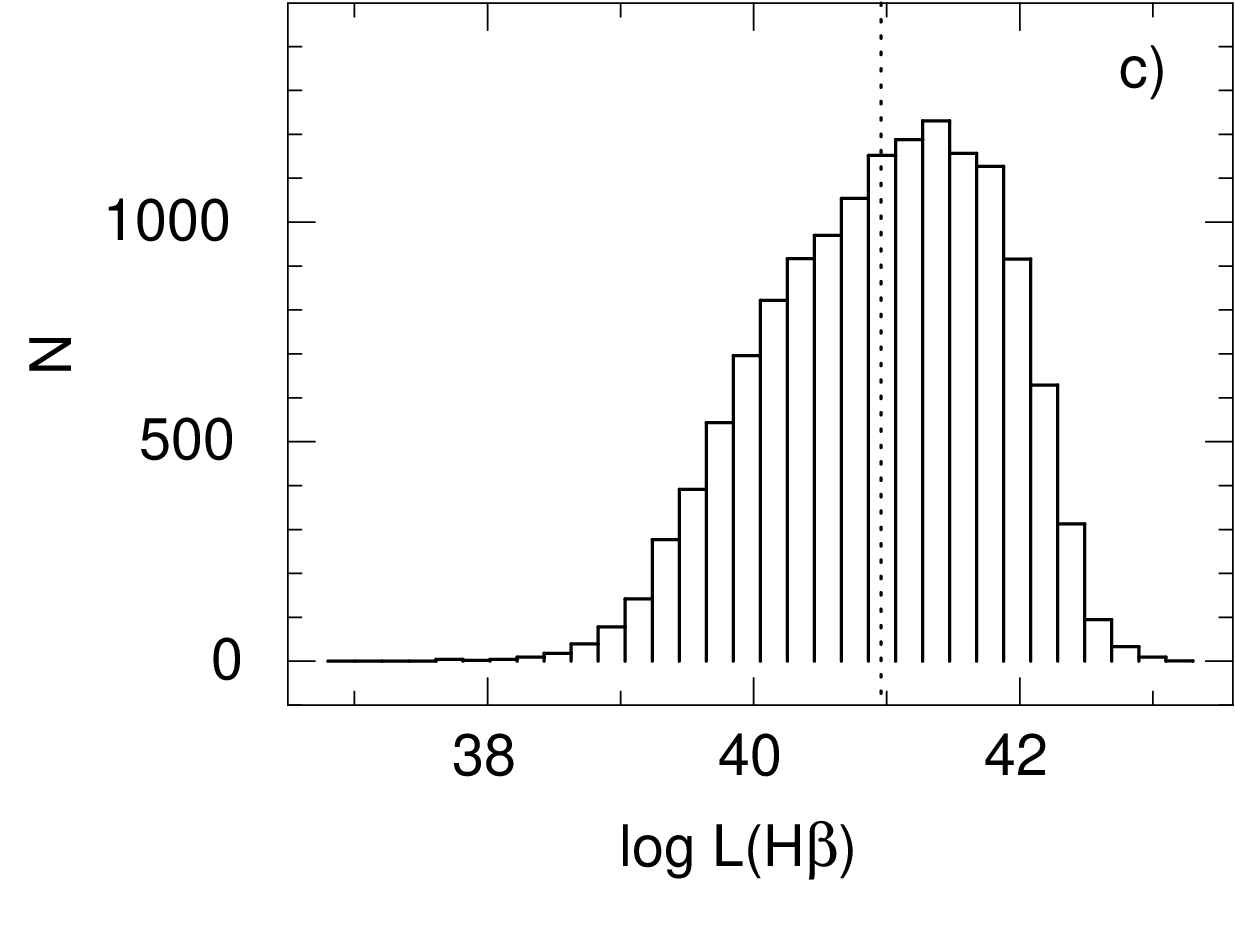}
\includegraphics[angle=-90,width=0.84\linewidth]{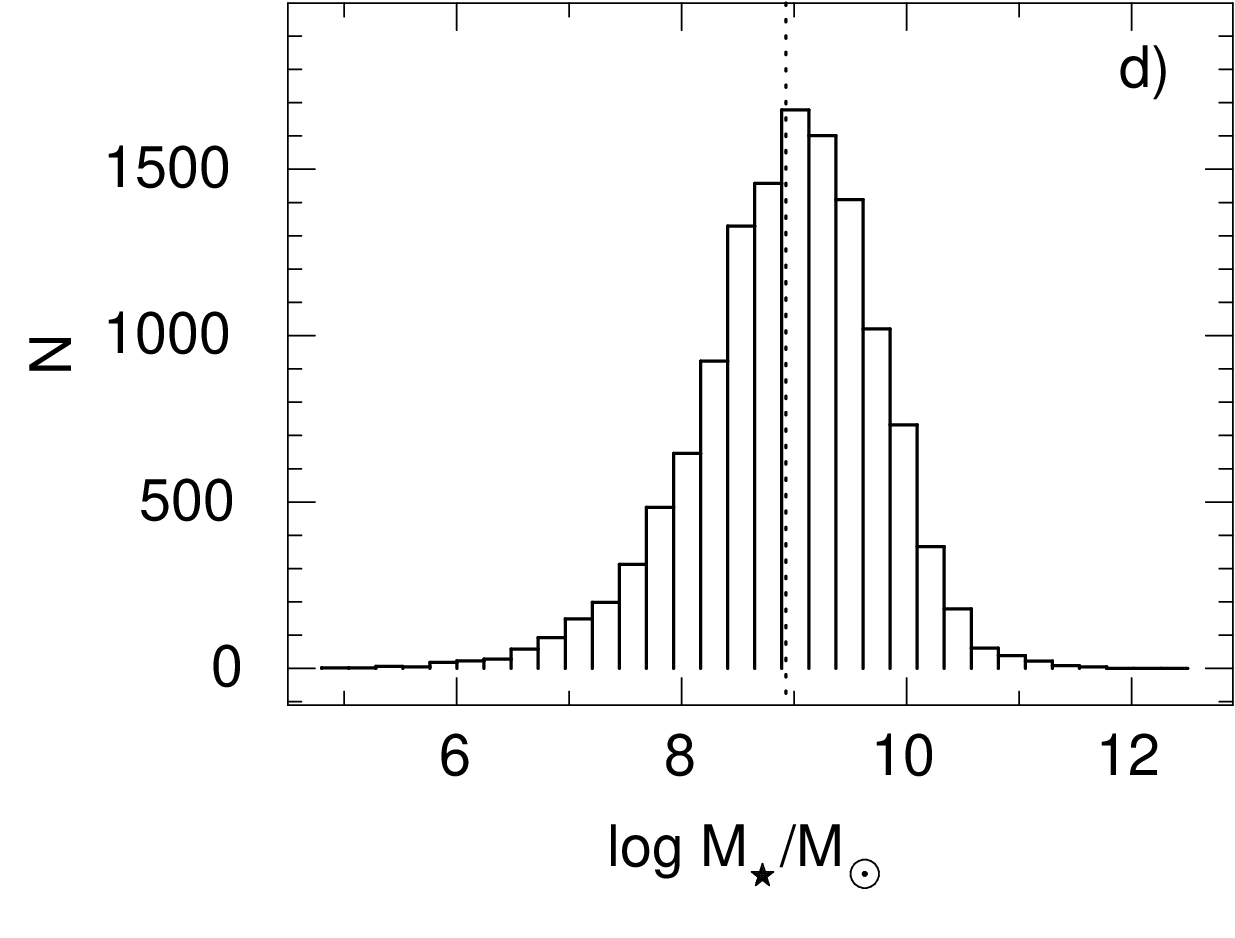}
\caption{Distributions of compact star-forming galaxies on (a) redshift $z$,
(b) rest-frame equivalent width EW(H$\beta$) of the H$\beta$ emission line,
(c) aperture- and extinction-corrected luminosity $L$(H$\beta$) of the 
H$\beta$ emission line expressed in erg s$^{-1}$, and (d) stellar galaxy mass
$M_\star$. Vertical dotted lines indicate average values.
\label{fig4}}
\end{figure}

\begin{figure}
\includegraphics[angle=-90,width=0.90\linewidth]{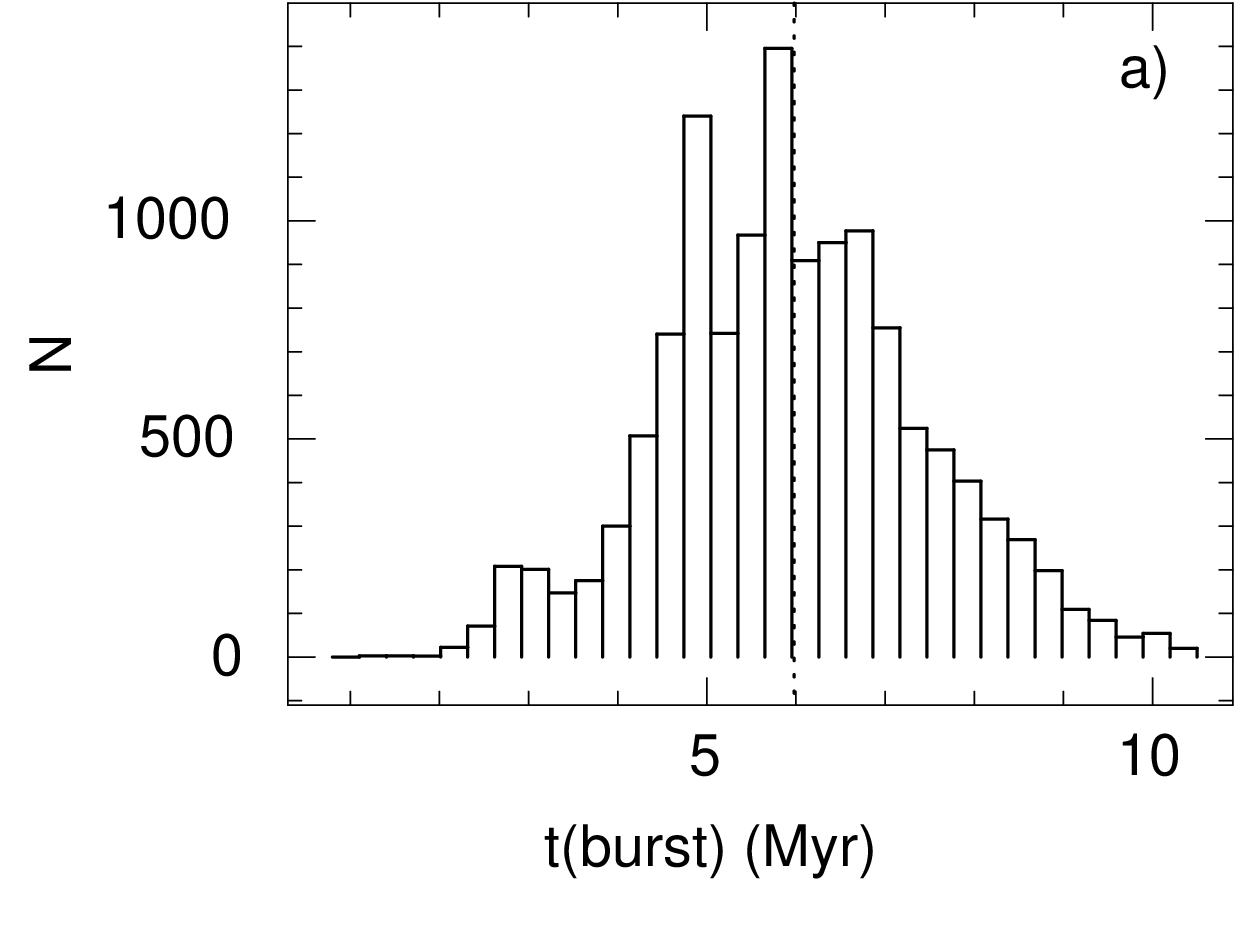}
\includegraphics[angle=-90,width=0.90\linewidth]{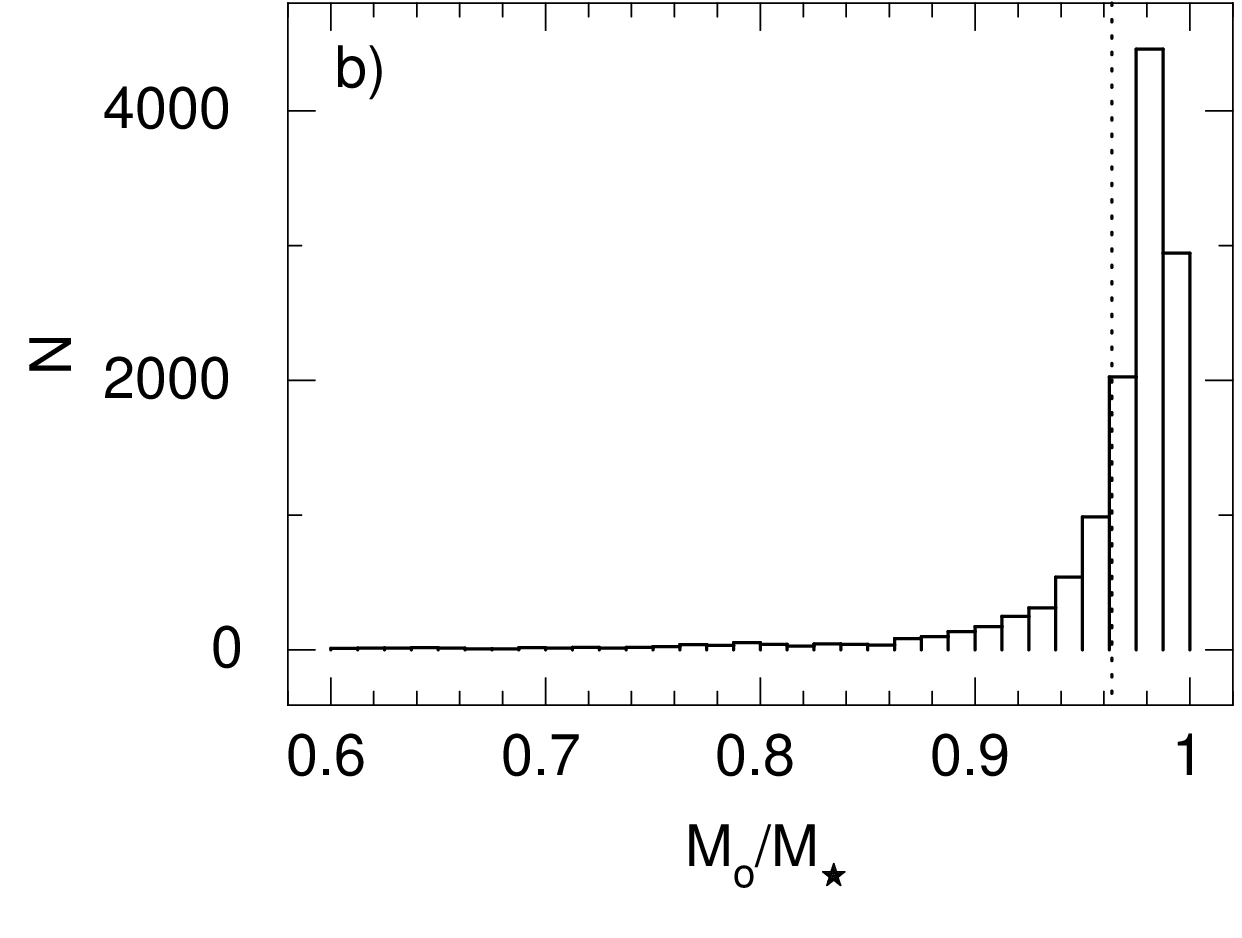}
\caption{Distributions of compact SFGs on (a) starburst 
ages and (b) the fraction of the mass of the old stellar
population. Vertical dotted lines indicate average values.
\label{fig5}}
\end{figure}

Selected galaxies are distributed over a redshift range of 0 -- 1 with average 
value $z$ $\sim$ 0.21 (Fig. \ref{fig4}a). The distributions of integrated 
characteristics of these galaxies EW(H$\beta$), $L$(H$\beta$), $M_\star$ are 
shown in Figs. \ref{fig4}b -- \ref{fig4}d. They closely resemble the 
properties of high-$z$ 
SFGs with strong emission lines \citep{I14a,I15}, making them the local 
counterparts of galaxies in the early universe.
\citet{I15} showed that low-$z$ compact SFGs follow the same 
mass - metallicity, luminosity - metallicity and mass - SFR 
relations with high-$z$ emission-line galaxies at $z$ = 2 -- 11 
studied by \citet{C14}, \citet{S14}, \citet{M08}, \citet{M14}, \citet{T14}, 
\citet{San15}, \citet{Sal15}, \citet{B12}, \citet{C13}, and \citet{Oe14}.
They have similar SFRs $\sim$ 1 -- 100 M$_\odot$ yr$^{-1}$ and
overlapping range of stellar masses $\sim$ 10$^{9}$ -- 10$^{11}$ M$_\odot$.
Furthermore, galaxies from our sample may 
contain a significant fraction of Lyman continuum leaking galaxies 
\citep{I16a,I16b}. 
Their high-$z$ counterparts likely were the main sources of the universe 
reionization at redshifts $\sim$ 5 -- 10. 

\section{The determination of integrated parameters}\label{sec3}

\subsection{Correction for extinction and aperture}

We measured emission-line fluxes and equivalent widths 
in the SDSS spectra using the IRAF\footnote {IRAF is the Image 
Reduction and Analysis Facility distributed by the National Optical Astronomy 
Observatory, which is operated by the Association of Universities for Research 
in Astronomy (AURA) under cooperative agreement with the National Science 
Foundation (NSF).} SPLOT routine. 
The observed decrement of several
hydrogen Balmer emission lines was used to correct the line fluxes relative to 
the H$\beta$ flux for two effects: (1) reddening adopting
the extinction curve of \citet{C89} and 
(2) underlying hydrogen stellar absorption that is derived simultaneously by an
iterative procedure as described in \citet{ITL94}.

The correction for reddening was done in two steps. First, the observed 
spectra, uncorrected for redshift, were corrected for the Milky Way 
extinction with $A(V)_{\rm MW}$ from 
the NASA Extragalactic Database (NED)\footnote{The NASA/IPAC Extragalactic
Database (NED) is operated by the Jet Propulsion Laboratory, California 
Institute of Technology, under contract with the National Aeronautics and 
Space Administration.} and $R(V)$ = $A(V)$/$E(B-V)$ = 3.1.
Then, the rest-frame spectra were corrected for the internal extinction 
$A(V)_{\rm int}$ of the galaxies, obtained from the hydrogen Balmer decrement 
after its correction for the Milky Way extinction.

The same $A(V)_{\rm MW}$, $A(V)_{\rm int}$, and $R(V)$
were used to correct {\sl GALEX} FUV and NUV fluxes.

\begin{figure}
\includegraphics[angle=-90,width=0.90\linewidth]{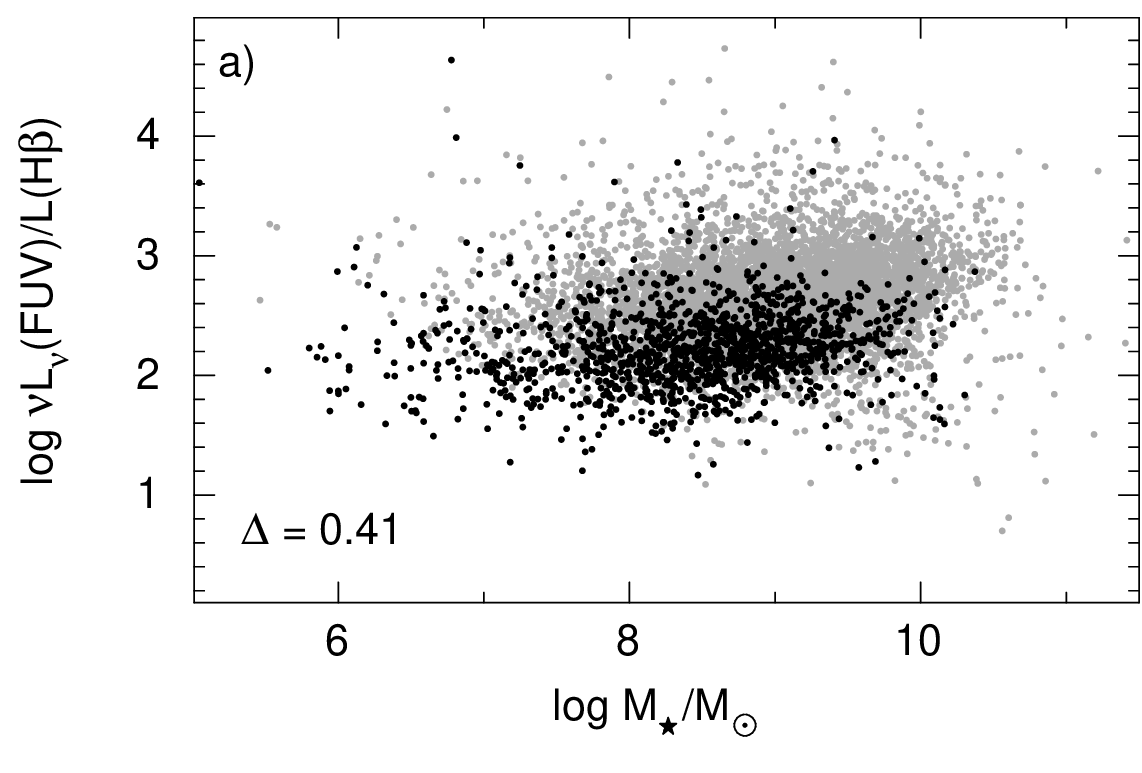}
\includegraphics[angle=-90,width=0.90\linewidth]{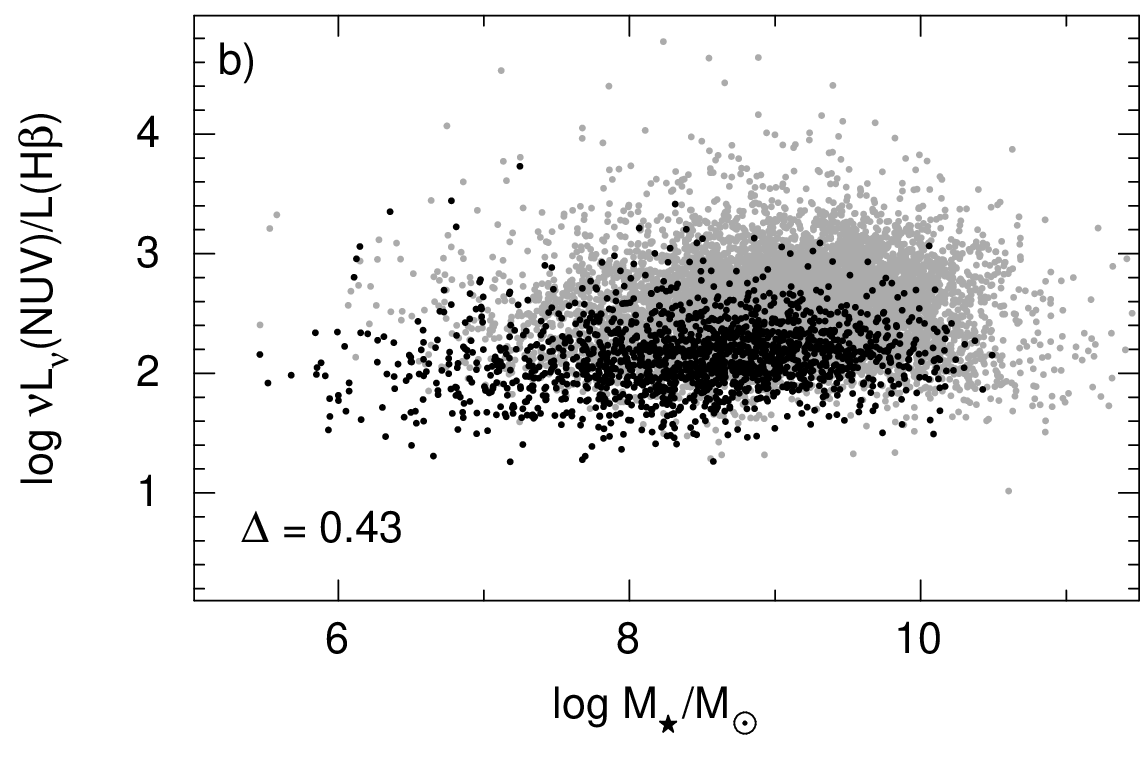}
\caption{The dependence of the extinction-corrected FUV-to-H$\beta$ (a)
and NUV-to-H$\beta$ (b) luminosity ratios on the stellar masses.
Additionally, $L$(H$\beta$) is corrected for the spectroscopic aperture.
Galaxies with rest-frame H$\beta$ equivalent widths 
EW(H$\beta$)~$<$ 100\AA\ and EW(H$\beta$)~$\geq$~100\AA\ are shown by 
grey and black dots, respectively. The values of $\Delta$ are the
differences in log scale between the averaged values of the FUV-to-H$\beta$ 
and NUV-to-H$\beta$ luminosity ratios for galaxies with low and high 
EW(H$\beta$)s.
\label{fig6}}
\end{figure}

In addition to extinction the H$\beta$ fluxes were also corrected for the 
aperture using the relation 2.5$^{r({\rm ap})-r}$, where $r$ and $r$(ap) are the 
SDSS $r$ total magnitude and the magnitude within the round spectroscopic 
aperture, respectively \citep{I15}. The aperture corrections of fluxes for 
most selected galaxies are small, less than a factor of 2 
(see Fig. \ref{fig2}b).
Only for less than 30\% of the sample a larger aperture correction is needed.

\subsection{H$\beta$ and UV luminosities and star formation rates}

The extinction- and aperture-corrected H$\beta$ fluxes were transformed to 
luminosities adopting luminosity distances derived with the cosmological 
calculator \citep[NED,][]{W06}, based on the cosmological 
parameters $H_0$=67.1 km s$^{-1}$Mpc$^{-1}$, $\Omega_\Lambda$=0.682, 
$\Omega_m$=0.318 \citep{P14}. Similarly, the {\sl GALEX} FUV and NUV 
luminosities were derived from the extinction-corrected fluxes. 

SFRs were obtained from the extinction- and aperture-corrected H$\beta$
luminosities adopting the \citet{K98} relation between SFR and $L$(H$\alpha$)
and an H$\alpha$/H$\beta$ flux ratio of 2.8.

\subsection{Stellar masses}

To fit the SEDs and to derive stellar masses we used the SDSS spectra
in the wavelength range $\sim$ 3800 -- 9200\AA\ for SDSS-II galaxies
and $\sim$ 3600 -- 10000\AA\ for SDSS-III galaxies,
and carried out a series of Monte Carlo simulations as described by
e.g. \citet{I15}. 

The star-formation history is approximated assuming a short burst with age 
$t$(burst) $<$ 10 Myr and a prior continuous star formation
with a constant SFR responsible for the older stars, with age starting
at $t_2$ = $t$(old) and ending at $t_1$ ($t_2 > t_1$ and zero time is now). 
Both $t_1$
and $t_2$ are varied in Monte Carlo simulations between 10 Myr and 15 Gyr.
The best solution of the SED is required to fulfill two conditions. First, only
models, in which EW(H$\beta$) agrees with the observed value within 5~\%, 
were selected. Second, the best modelled SED among selected models for each 
set of fixed parameters was found from $\chi ^2$ minimization of 
the deviation between the modelled and the observed continua.

The best solutions can be found for different combinations of evolutionary 
tracks, stellar atmosphere models and initial mass functions
with the typical uncertainties of $\sim$ 0.2 dex for the stellar mass. 
In Fig. \ref{fig5}a we show the distribution of burst ages $t$(burst) with
the maximum of $\sim$ 6 Myr, corresponding to the equivalent width 
EW(H$\beta$) $\sim$ 30\AA, while the burst age of 10 Myr corresponds to
EW(H$\beta$) $\sim$ 7\AA. The mass fraction of the old stellar population
formed continuously with a constant SFR between $t_2$ and $t_1$ is shown in
Fig. \ref{fig5}b with an average value of $\sim$ 96\% corresponding to the 
mass ratio of the old-to-young stellar population of $\sim$ 24.

We note that the photometric data were not used in the SED 
fitting, but they are in general consistent with the SEDs derived 
from the SDSS optical spectra \citep[e.g. ][]{I14a}. In particular, 
we do not use the near-infrared (NIR) photometric data, which are known only
for less than 50\% of galaxies from our sample. It is often assumed that
emission of a galaxy in the NIR is produced by the old stellar
population. This assumption may not be true for SFGs. The 
luminosity ratio of the 3 Myr and 10 Gyr instantaneous bursts with equal
masses at 10000\AA\ is $\sim$ 25. Therefore, luminosities at this wavelength
are equal if the mass of old stellar population is 25 times higher than that
of the young stellar population, very similar to the average value 
in Fig. \ref{fig5}b. This means that, on average, the contribution of the
old stellar population in the red part of the SFG SDSS spectrum is
considerable, allowing for a reliable mass determination of the old population.
However, the contribution of the young population is not negligible.
The situation is more complex for the SFGs with EW(H$\beta$) $>$ 100\AA\ 
because the contribution of the nebular continuum in the visible and NIR ranges
is high and it needs to be taken into account \citep{I11}.
On the other hand, the average contribution of the old stellar population to
the continuum near the H$\beta$ emission line is only $\sim$ 4 -- 5\%, 
indicating that EW(H$\beta$) is a reliable tracer of the starburst 
age in our compact SFGs.

\section{Bursting vs continuous star formation}\label{sec4}

In Fig. \ref{fig6} we show the relations between the FUV-to-H$\beta$
and NUV-to-H$\beta$ luminosity ratios and the stellar masses
for the entire sample. We split the sample into objects with high H$\beta$
equivalent widths EW(H$\beta$) $\geq$ 100\AA\ (black dots) and with low H$\beta$
equivalent widths EW(H$\beta$) $<$ 100\AA\ (grey dots). Clear offsets between
the two sets of data are present with ratios, which are on average
by $\Delta$ $\sim$ 0.4 dex higher for galaxies with low EW(H$\beta$)s.

Dispersions $\sim$ 0.3 -- 0.4 dex of galaxies in Fig. \ref{fig6}
are caused in part by uncertainties of the {\sl GALEX} FUV and NUV magnitudes,
which for our faint galaxies are as high as 0.3 -- 0.5 mag, corresponding to
the FUV and NUV luminosity uncertainties $\sim$ 0.1 -- 0.2 dex. Another
source of the dispersions in Fig. \ref{fig6} are uncertainties of the extinction
determination from the hydrogen Balmer decrement, which typically correspond
to the uncertainties $\sim$ 0.1 -- 0.2 mag in $A(V)$. These uncertainties
translate to the uncertainties $\sim$ 0.5 -- 1.0 mag in the FUV and NUV
ranges, corresponding to uncertainties $\sim$ 0.2 -- 0.4 dex in the FUV and
NUV luminosities. These two sources of statistical uncertainties are
sufficient to explain the observed dispersions in Fig. \ref{fig6}. However,
they can not explain the offsets in distributions of SFGs with low and high
EW(H$\beta$)s.

\begin{figure}
\hbox{
\includegraphics[angle=-90,width=0.90\linewidth]{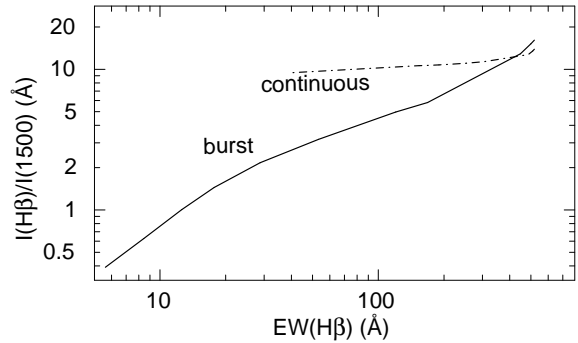}
}
\caption{
The dependence of the H$\beta$ to monochromatic 1500\AA\ flux ratio on the 
H$\beta$ equivalent width modelled with the Starburst99 
code \citep{L99,L14}. The instantaneous burst and continuous star formation 
with a constant star formation rate are shown by solid and dash-dotted lines, 
respectively.
\label{fig7}}
\end{figure}

\begin{figure}
\includegraphics[angle=-90,width=0.90\linewidth]{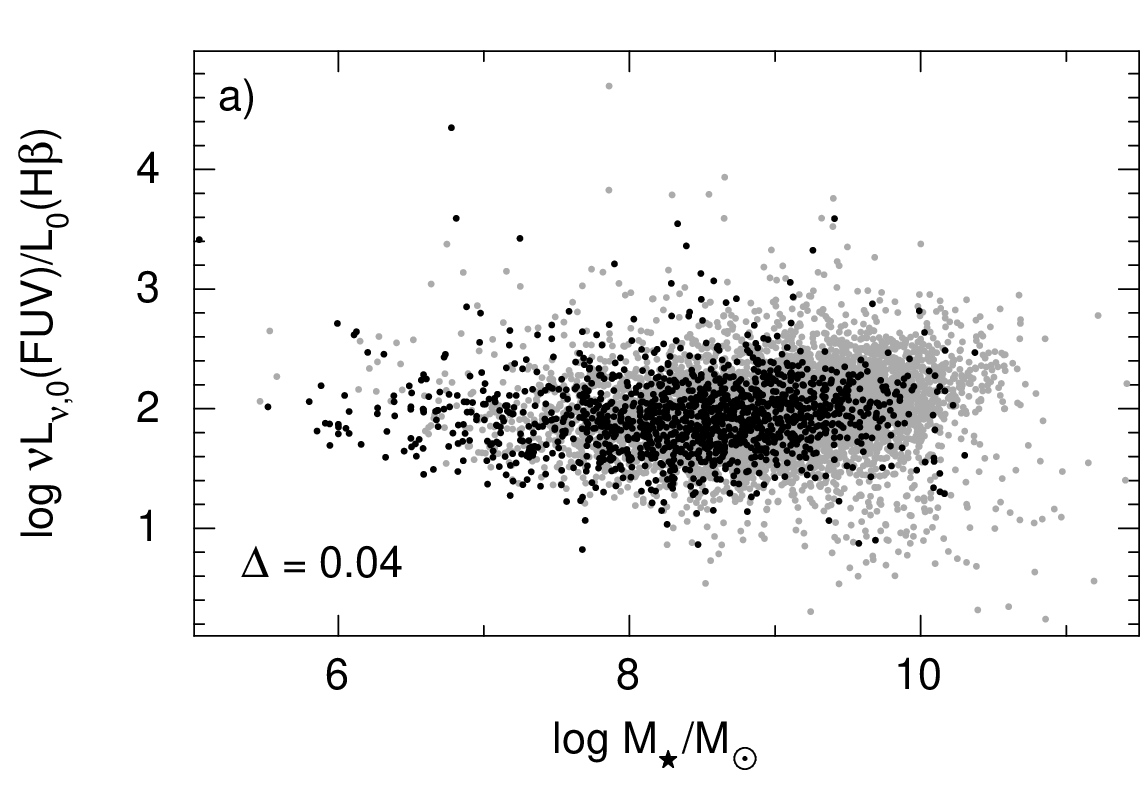}
\includegraphics[angle=-90,width=0.90\linewidth]{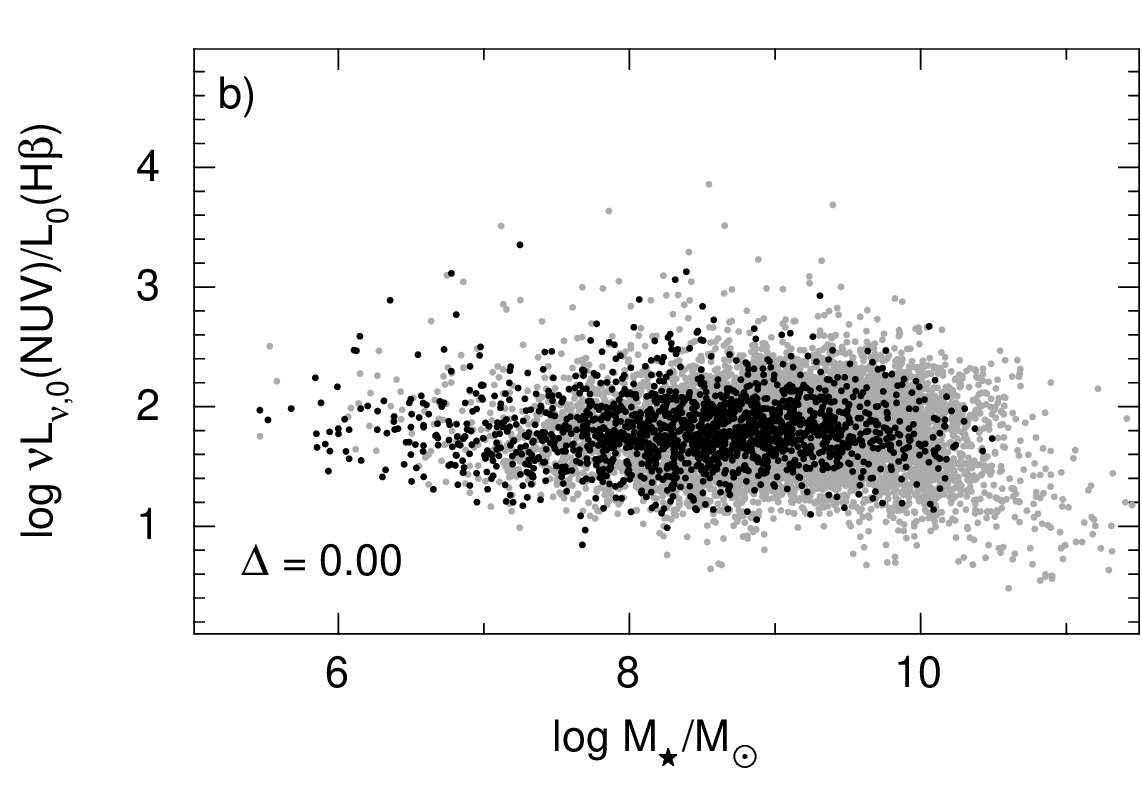}
\caption{Same as in Fig. \ref{fig6}, but H$\beta$, FUV and NUV luminosities 
are reduced to a zero starburst age.
\label{fig8}}
\end{figure}

One of the possible explanations of these offsets is the bursting nature
of star formation in the galaxies in the entire range of stellar masses
meaning that star bursts dominate the H$\beta$ and UV 
radiation even in the most massive galaxies from our sample. 
To study this possibility we consider temporal variations of 
the FUV and H$\beta$ luminosities for bursting and continuous star formation
using Starburst99 models \citep{L99,L14}. In Fig. \ref{fig7} we show the
dependence of the H$\beta$-to-FUV luminosity ratio on the H$\beta$ equivalent
width EW(H$\beta$) for the instantaneous burst (solid line) and continuous
star formation (dash-dotted line). Here we have adopted a Salpeter IMF 
\citep{S55} with the upper mass limit of 100 M$_\odot$ and the low mass limit of
0.1 M$_\odot$, Geneva evolutionary tracks \citep{M94} of
non-rotating stars and a combination of stellar atmosphere
models \citep{L97,S92}. It is seen that the 
H$\beta$-to-FUV luminosity ratio is slowly decreasing with decreasing
EW(H$\beta$) in the case of the scenario with continuous star formation
while a steep decrease is found for the burst model.

The EW(H$\beta$) is a measure of the age in the instantaneous burst model
with the higher value for the younger age. 
We use EW(H$\beta$) 
and Starburst99 models to reduce H$\beta$ luminosity to zero age in the
burst model:
\begin{equation}
\Delta \log L({\rm H}\beta) = 2.7 - \log [{\rm EW}({\rm H}\beta)]  \label{eq:Hbcorr}
\end{equation}
for $\log {\rm EW}({\rm H}\beta) \leq 2.7$, otherwise 
$\Delta \log L({\rm H}\beta) = 0$.

We note that the age correction is needed not only for the H$\beta$ luminosity
but also for the FUV and NUV luminosities which are produced by young stars.
In the bursting model, FUV and NUV luminosities decrease more slowly than
the H$\beta$ luminosity because they are produced by longer-lived stars.

Using the same instantaneous burst models obtained with the Starburst99 code 
we obtain the following equations for reduction of FUV and NUV luminosities to 
zero age:
\begin{equation}
\Delta \log L({\rm FUV}) = 0.39 \times \Delta \log L({\rm H}\beta), \label{eq:FUVcorr}
\end{equation}
\begin{equation}
\Delta \log L({\rm NUV}) = 0.30 \times \Delta \log L({\rm H}\beta). \label{eq:NUVcorr}
\end{equation}

The dependencies of the FUV-to-H$\beta$ and NUV-to-H$\beta$ luminosity ratios 
on the stellar masses are shown in Fig.
\ref{fig8}, where all luminosities are reduced to a zero age. It is seen that
no offsets are present between the galaxies with low and high EW(H$\beta$)s.
The differences $\Delta$ between average FUV-to-H$\beta$ and NUV-to-H$\beta$ 
luminosity ratios for galaxies with high and low EW(H$\beta$)s are $\approx$ 0.

Instead, offsets remain in the
case of continuous star formation because FUV-to-H$\beta$ and NUV-to-H$\beta$
luminosity ratios vary little depending on the EW(H$\beta$) or equivalently
on the age (dash-dotted line in Fig. \ref{fig7}). Therefore, we conclude that
stellar populations producing H$\beta$ and UV radiation 
in compact SFGs from our sample are formed during recent 
short bursts. Of course, we cannot exclude the continuous star formation in
our galaxies. However, even if present, its contribution to the galaxy
H$\beta$ emission line and UV continuum luminosities must be very small. 

\section{Relations between global parameters, corrected for the burst age}
\label{sec5}

Our finding emphasizes
the importance of the age correction in order to compare properties of
star formation in similar conditions. Such correction would greatly reduce
scatter of points on various diagrams including H$\beta$ and UV luminosities
and SFRs obtained from these luminosities.

\subsection{H$\beta$ luminosity and star formation rate}

We first consider the relation between the H$\beta$ luminosity and the stellar
mass. It can be transformed to the relation between the SFR and
stellar mass by using the \citet{K98} equation.

\begin{figure}
\includegraphics[angle=-90,width=0.90\linewidth]{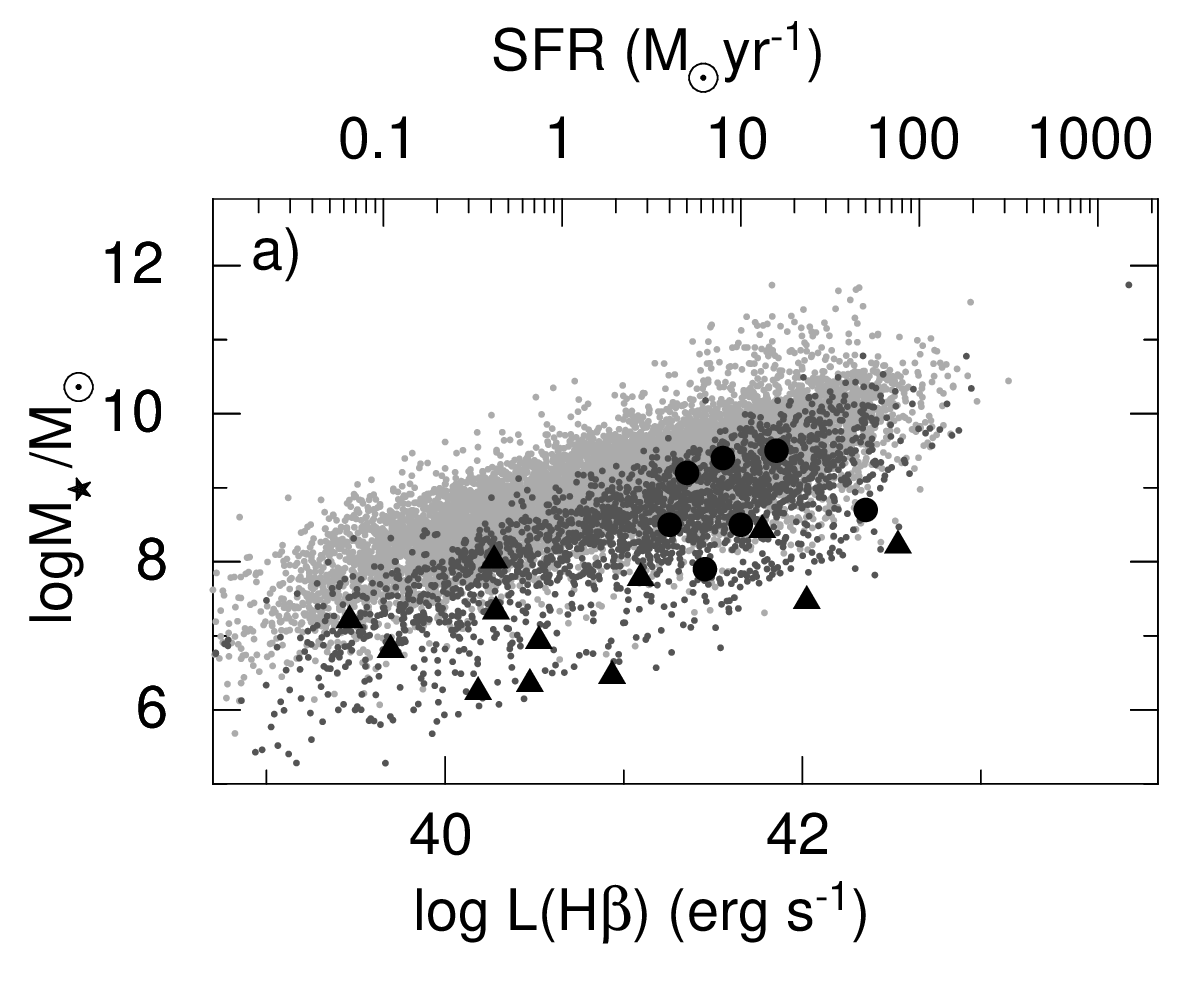}
\includegraphics[angle=-90,width=0.90\linewidth]{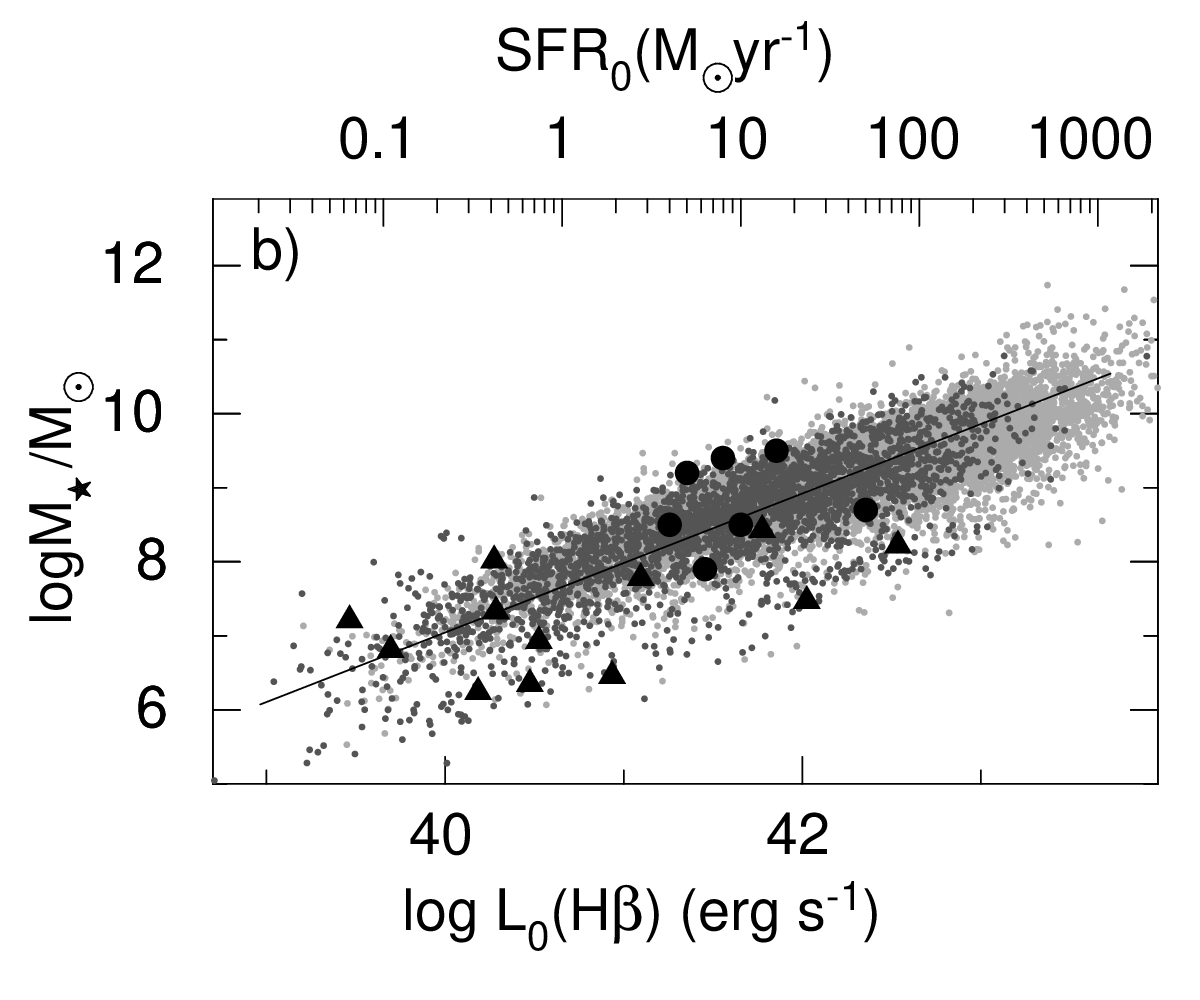}
\caption{(a) The dependence of the stellar mass $M_\star$ on the extinction- and
aperture-corrected H$\beta$ luminosity $L$(H$\beta$) (lower axis) and
star formation rate SFR (upper axis). (b) Same as in (a) but 
$L$(H$\beta$) and SFR are reduced to zero burst age. In both panels, galaxies 
with rest-frame H$\beta$ equivalent widths EW(H$\beta$) $<$ 100\AA\ and
EW(H$\beta$) $\geq$ 100\AA\ are shown by light-grey and dark-grey dots, 
respectively. The linear regression in (b) for the entire sample
is shown by a black solid line. For comparison, the
high-redshift galaxy candidates at $z$ = 9 -- 11 \citep{C13,Oe14} 
and low-mass Ly$\alpha$ emitting galaxies at $z$ = 3 -- 6 \citep{K16}
are shown by filled circles and filled triangles, respectively.
\label{fig9}}
\end{figure}

In Fig. \ref{fig9}a we show the stellar mass -- H$\beta$ luminosity diagram
for the galaxies from our sample of compact SFGs in the case
when $L$(H$\beta$) is not corrected for the burst age. 
On the upper abscissa we also show the SFR scale. 
As expected from the
above discussion, a clear offset is present for the galaxies with low and 
high EW(H$\beta$)s shown by light-grey and dark-grey dots, respectively, implying
that corrections for the starburst age are needed to produce the unbiased 
relation. For comparison, we show
high-$z$ galaxy candidates at $z$ = 9-11 \citep{C13,Oe14} (filled circles), 
which are considered as the ones among the first galaxies formed in the 
universe, and the low-mass Ly$\alpha$ emitting galaxies at $z$ = 3-6
\citep{K16} (filled triangles). It is seen that
the location of these high-$z$ galaxies is in good agreement with the location
of low-$z$ SFGs from our sample with high EW(H$\beta$) $\geq$ 100\AA\ 
(dark-grey dots), indicating similar properties.
We note that no correction for the burst age using the H$\beta$ emission line is
possible for the high-$z$ galaxies because EW(H$\beta$)s were not presented
by \citet{C13}, \citet{Oe14} and \citet{K16}.

\begin{figure}
\includegraphics[angle=-90,width=0.90\linewidth]{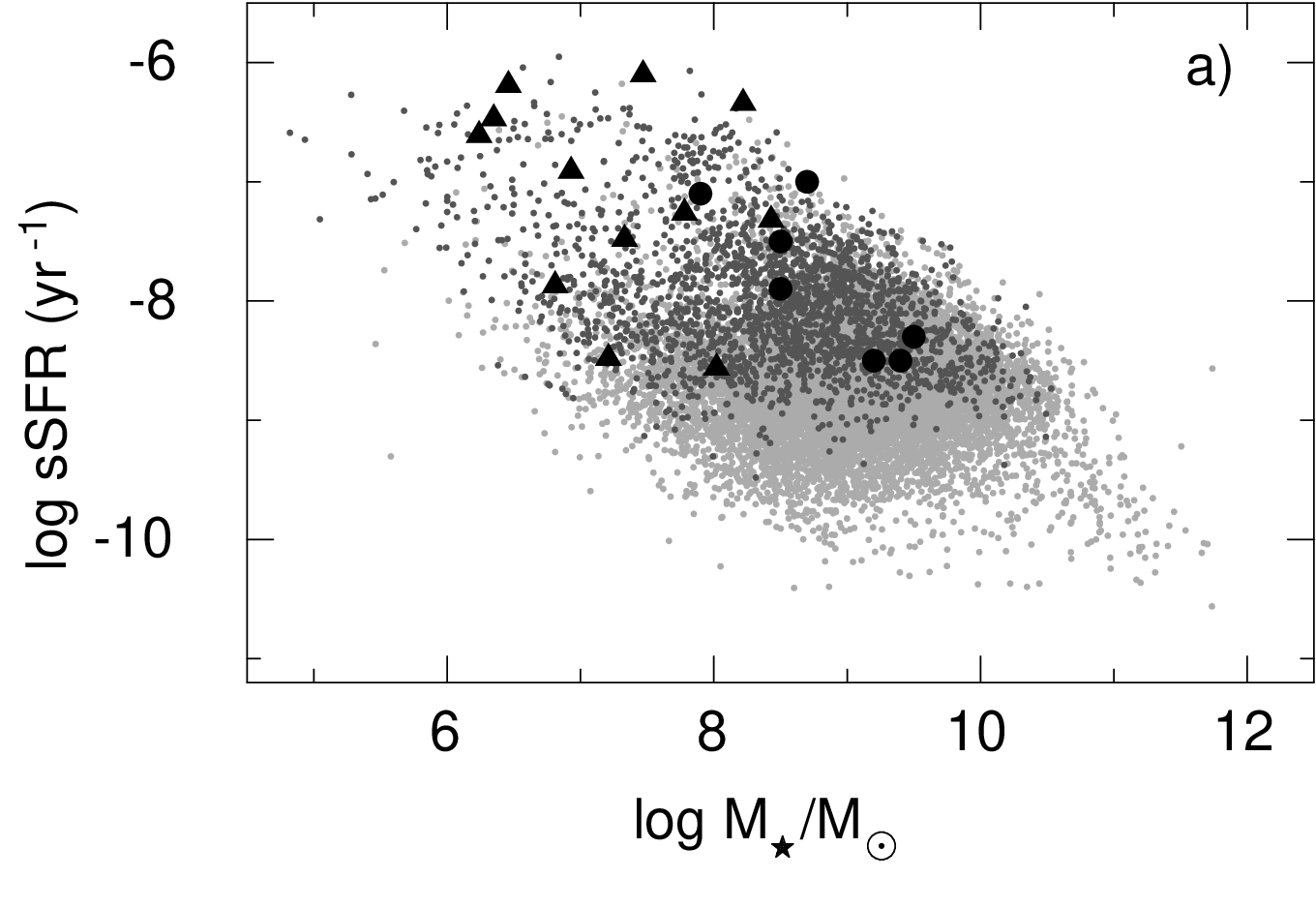}
\includegraphics[angle=-90,width=0.90\linewidth]{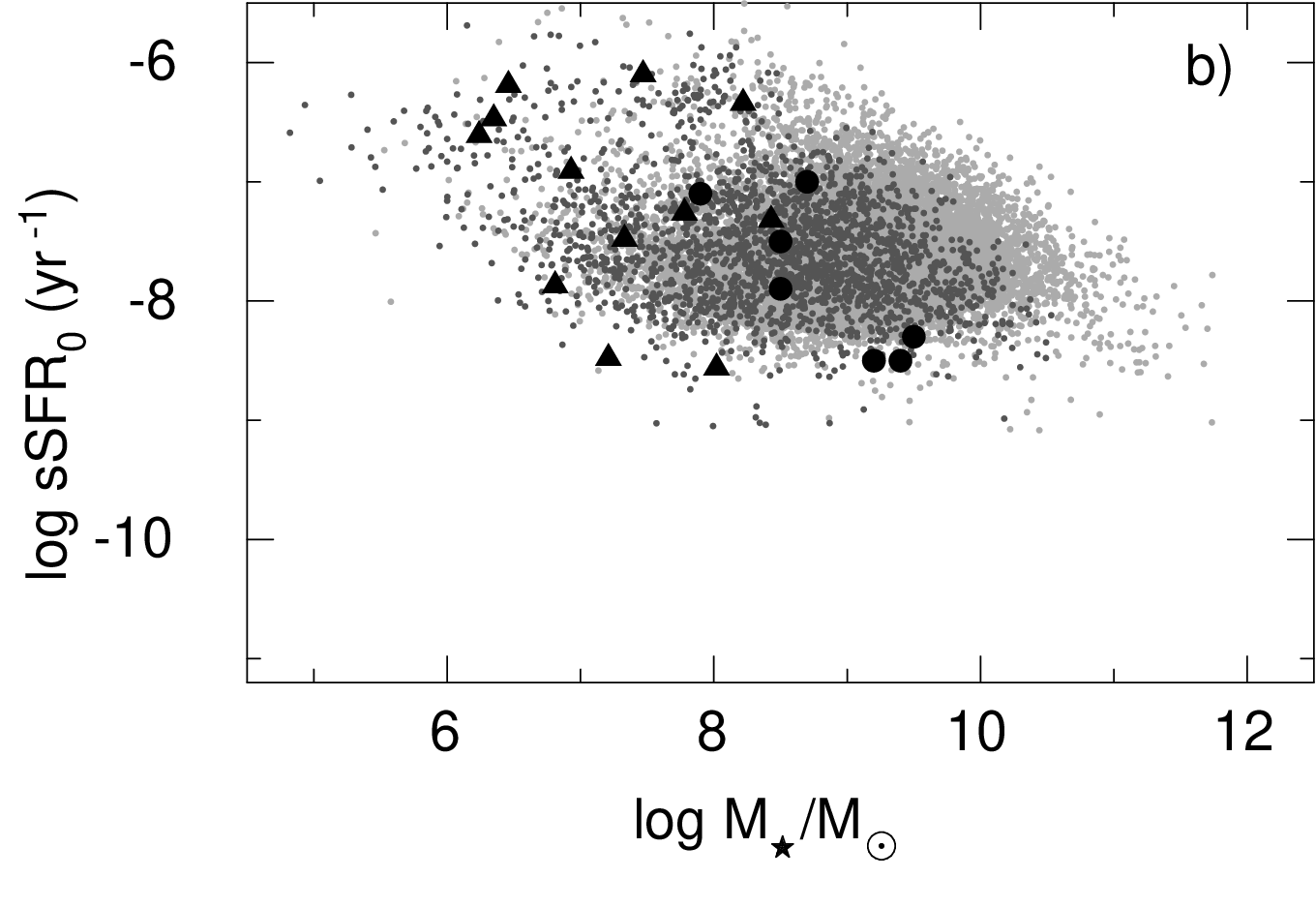}
\caption{The dependence of the extinction- and aperture-corrected specific
star formation rate sSFR on the stellar mass $M_\star$. (b) Same as in (a) but 
sSFR is reduced to zero burst age. In both panels symbols are the same as in
Fig. \ref{fig9}.
\label{fig10}}
\end{figure}

We show in Fig. \ref{fig9}b the same stellar mass -- H$\beta$ luminosity 
relation as in Fig. \ref{fig9}a but the $L$(H$\beta$) is reduced to zero age.
We denote the corrected H$\beta$ luminosity as $L_0$(H$\beta$) and the 
respective star formation rate as SFR$_0$. It is seen that no offset is present
between the galaxies with low and high EW(H$\beta$)s and the relation is
much tighter as compared to that in Fig. \ref{fig9}a. This relation
can be approximated by the dependence:
\begin{equation}
\frac{M_\star}{{\rm M}_\odot} = 3.08 \times 10^{-31}L({\rm H}\beta)^{0.94}
= 5.06 \times 10^7 {\rm SFR}^{0.94}, 
\label{MSFR}
\end{equation}
where $L$(H$\beta$) is in erg s$^{-1}$ and SFR in M$_\odot$ yr$^{-1}$. 
It is very close to a linear
dependence suggesting that the SFR is nearly proportional to the stellar mass.
We also note that the distribution of low-$z$ compact SFGs is coincident with
the distribution of high-$z$ SFGs shown by filled circles and
triangles despite the fact that no age correction is applied for high-$z$ 
galaxies. However, SFRs for high-$z$ galaxies are derived not from the H$\beta$
emission line but from the UV continuum luminosities, for which age corrections
are much smaller. Furthermore, distributions of low-$z$ SFGs with high
EW(H$\beta$)s and high-$z$ SFGs in Fig. \ref{fig9}a are very similar  
indicating that the emission of high-$z$ galaxies is also dominated by 
young bursts. Overall, we assume that burst age corrections for high-$z$ 
galaxies shown in Fig. \ref{fig9} should be small.

\subsection{Specific star formation rate}

The offset of the specific star formation rate 
sSFR $\equiv$ SFR/($M_\star$/M$_\odot$) is clearly seen between the galaxies with 
low and high EW(H$\beta$)s resulting in a broad distribution of the compact 
SFGs on the sSFR -- $M_\star$ diagram (Fig. \ref{fig10}a).

On the other hand, the age correction results in a much narrower distribution
(Fig. \ref{fig10}b). This distribution is almost independent on $M_\star$,
as expected, because the SFR is almost linearly 
increased with increasing $M_\star$ (Eq. \ref{MSFR}). We also note that
the very high values of sSFR$_0$ of $\sim$ 10 -- 100 Gyr$^{-1}$, among the
highest known so far for galaxies, imply very efficient conversion
of the interstellar matter to stars. These values are similar to the ones 
derived for the highest-redshift galaxies shown in Fig. \ref{fig10} by
filled circles and triangles. However, due to the bursting nature
of star formation in our low-$z$ compact and high-$z$ SFGs, the specific 
star formation rate averaged over large time intervals of 1 -- 10 Gyr 
would be 
lower. This averaged value is uncertain because it depends
both on the time interval adopted for averaging and the unknown number of
starbursts during the same interval. It is likely that star formation
in our compact galaxies converts interstellar matter to the stellar mass on 
time scales which may be considerably lower than the cosmological time.

\section{Conclusions}\label{sec6}

In this paper we present a study of the integrated characteristics of
a sample of $\sim$ 14000 compact star-forming galaxies (SFGs) at redshifts
$z$ $\sim$ 0 -- 1 selected from the Data Release 12 of the
Sloan Digital Sky Survey. Our main results are as follows.

1. We find that the star formation in our galaxies occurs in short strong
bursts which dominate the H$\beta$ and UV continuum radiation while
the contribution of the continuous star formation to the H$\beta$ and UV 
luminosities, even if present, is very low.

2. Since the H$\beta$ and UV luminosities of the compact SFGs
are rapidly changing with time, the correction for the starburst age is needed
to produce homogeneous sets of data in different samples at various redshifts
for their comparison.

3. We find that the scatter of points in the relation stellar mass $M_\star$ -- 
star
formation rate (SFR) is greatly reduced if the H$\beta$ luminosities and SFRs 
are reduced to zero starburst age. It is shown that SFRs and $L$(H$\beta$)s are
almost linearly increasing with $M_\star$. The scatter of specific 
star formation rates (sSFRs) is also greatly reduced if the correction for the 
burst age is taken into account. We find that sSFRs for our sample galaxies 
corrected for the starburst age do not
depend on $M_\star$, which is a consequence of a nearly linear dependence
between $M_\star$ and SFR, and are among the highest known for SFGs.

4. It is shown that average stellar masses, H$\beta$ luminosities, SFRs and
sSFRs of our galaxies are similar
to the respective characteristics of highest-redshift SFGs
and imply that low-$z$ compact SFGs may be considered as 
local counterparts of the galaxies in the early universe.

\section*{Acknowledgements}

Funding for the SDSS and SDSS-II was provided by the Alfred P. Sloan 
Foundation, the Participating Institutions, the National Science Foundation, 
the U.S. Department of Energy, the National Aeronautics and Space 
Administration, the Japanese Monbukagakusho, the Max Planck Society, and the 
Higher Education Funding Council for England. 
Funding for SDSS-III has been provided by the Alfred P. Sloan Foundation, 
the Participating Institutions, the National Science Foundation, and the U.S. 
Department of Energy Office of Science. The SDSS-III web site is 
http://www.sdss3.org/. SDSS-III is managed by the Astrophysical Research 
Consortium for the Participating Institutions of the SDSS-III Collaboration. 
GALEX is a NASA mission  managed  by  the  Jet  Propulsion  Laboratory.
This research has made use of the NASA/IPAC Extragalactic Database (NED) which 
is operated by the Jet  Propulsion  Laboratory,  California  Institute  of  
Technology,  under  contract with the National Aeronautics and Space 
Administration.






\bsp	
\label{lastpage}
\end{document}